\pdfoutput=1

\documentclass[11pt]{article}

\usepackage[]{emnlp2021}

\usepackage{times}
\usepackage{latexsym}

\usepackage[T1]{fontenc}

\usepackage[utf8]{inputenc}

\usepackage{microtype}

\usepackage{microtype}
\usepackage{graphicx} 
\usepackage{makecell}
\usepackage{amsmath}
\usepackage[ruled,linesnumbered]{algorithm2e}

\usepackage{subcaption}
\usepackage{multirow}

\title{Efficient-FedRec: Efficient Federated Learning Framework\\for Privacy-Preserving News Recommendation}

\author{
  Jingwei Yi$^1$, Fangzhao Wu$^2$, Chuhan Wu$^3$, Ruixuan Liu$^4$, Guangzhong Sun$^1$, Xing Xie$^2$\\
  $^1$University of Science and Technology of China
  $^2$Microsoft Research Asia\\
  $^3$Tsinghua University
  $^4$Renmin University of China \\
  {\tt yjw1029@mail.ustc.edu.cn} {\tt \{wufangzhao,wuchuhan15\}@gmail.com} \\
  {\tt ruixuan.liu@ruc.edu.cn} \\
  {\tt gzsun@ustc.edu.cn xingx@microsoft.com}
}

\begin{document}
\maketitle
\begin{abstract}
News recommendation is critical for personalized news access. 
Most existing news recommendation methods rely on centralized storage of users' historical news click behavior data, which may lead to privacy concerns and hazards. 
Federated Learning is a privacy-preserving framework for multiple clients to collaboratively train models without sharing their private data. 
However, the computation and communication cost of directly learning many existing news recommendation models in a federated way are unacceptable for user clients. 
In this paper, we propose an efficient federated learning framework for privacy-preserving news recommendation.
Instead of training and communicating the whole model, we decompose the news recommendation model into a large news model maintained in the server and a light-weight user model shared on both server and clients, where news representations and user model are communicated between server and clients.
More specifically, the clients request the user model and news representations from the server, and send their locally computed gradients to the server for aggregation.
The server updates its global user model with the aggregated gradients, and further updates its news model to infer updated news representations.
Since the local gradients may contain private information, we propose a secure aggregation method to aggregate gradients in a privacy-preserving way. 
Experiments on two real-world datasets show that our method can reduce the computation and communication cost on clients while keep promising model performance.

\end{abstract} 
\section{Introduction}
With the explosion of online information, the large quantities of news generated every day may overwhelm users and make them difficult to find the news they are interested in.
To tackle this problem, many news recommendation methods \cite{an-etal-2019-neural, wang-etal-2020-fine, wu-etal-2019-neural-news-recommendation,qi-etal-2021-hierec} have been proposed to display news according to users' personalized interests.
These methods are usually composed of two core modules, i.e., user model and news model.
The user model is used to learn user representations from user historical click behaviors.
For example,~\citet{wang2018dkn} use a candidate-aware attention network as the user model to help capture user interests in candidate news. 
The news model is used to learn news representations from news content.
For example,~\citet{wu2019neural} apply multi-head self attention network to capture the interactions between words in news model.
With the success of pre-trained language models (PLM) in NLP, a few PLM-empowered news recommendation methods have been proposed and achieve remarkable performance.
For example,~\citet{wu2021empowering} apply pre-trained language models to enhance news modeling. 
However, these methods require centralized storage of user behaviors, which are highly privacy-sensitive~\cite{shin2018privacy}.
Collecting private user data has raised many concerns~\cite{wu2019tamf}.
Moreover, due to the adoption of some data protection regulations such as GDPR\footnote{https://gdpr-info.eu}, it might not be able to analyze centralized user data in the future.

Federated learning~\cite{mcmahan2017communication} is a privacy-preserving method to train models on  the private data decentralized on a large number of clients.
In federated learning, each user keeps a local copy of model, and compute local model gradients with their local private data.
A central server coordinates the clients and aggregates local gradients to update the global model.
Recently,~\citet{qi-etal-2020-privacy} proposed a FedRec method to train news recommendation models using federated learning.
However, the model sizes of many existing news recommendation methods are large, especially their news models.
For example, PLM-NR~\cite{wu2021empowering} has 110.7M parameter in total, 110M of which are in the news model (BERT-Base version).
Thus, the communication and computation costs of FedRec can be too high for clients with rather limited computation resource.

In this paper, we propose an efficient federated learning framework for privacy-preserving news recommendation named Efficient-FedRec\footnote{https://github.com/yjw1029/Efficient-FedRec}.
In our framework, we decompose the news recommendation model into a large news model and a light-weight user model.
Instead of training and communicating the whole model, in our approach the clients only request the user model and the representations of news involved in their local behaviors from the server.
The clients locally compute the gradients of the user model and news representations on their local data, and send them to the server for aggregation.
The central server uses the aggregated user model gradients  to update its maintained global user model, and update the news model based on the aggregated news representation gradients.
The updated news model is further used to infer updated news representations.
The above process is repeated for multiple rounds until the model gets converges.
In order to protect user privacy in model training, we develop a secure aggregation protocol based on the multi-party computation framework for privacy-preserving gradient aggregation.
We exchange the news representations in the union news set involved by a group of user behaviors to protect the click history of a specific user.
We conduct plenty of experiments on two real-world datasets and the results show that our approach can effectively reduce the computation and communication cost on clients for federated news recommendation model training.

The main contributions of this work include:
\begin{itemize}
    \setlength\itemsep{0.1em}
    \item We propose an efficient federated learning framework for privacy-preserving news recommendation, which can effectively reduce the computation and communication cost on the user side.
    \item We develop an effective and efficient secure aggregation protocol to protect user privacy in model training.
    \item We conduct thorough experiments on two real-world datasets to verify the effectiveness and efficiency of our approach.
\end{itemize}

\section{Related Works}

\subsection{Neural News Recommendation}
Personalized news recommendation is an important technique to alleviate the information overloading problem and improve user reading experience.
Many deep learning based recommendation methods have been proposed ~\cite{wu2019npa,okura2017embedding,zhu2019dan,qi-etal-2021-pp,10.1145/3404835.3462861}.
They usually contain two core modules, i.e., user model and news model.
For example, ~\citet{an-etal-2019-neural} propose to use a CNN to learn contextual word embedding and an attention layer to select informative words.
They combine long-term interests and short-term interests of users by using user id embeddings and a GRU network in user model.
These methods learn news representations based on shallow NLP models, which is hard to well capture the news semantic information.
Recently, pre-trained language models (PLM) achieve great success in NLP~\cite{Devlin2019BERTPO,liu2019roberta,bao2020unilmv2}.
A few PLM-empowered news recommendation methods have been proposed.
For example, ~\citet{wu2021empowering} propose PLM-NR to empower news modeling by applying pre-trained language.
They replace the news encoder in previous methods with pre-trained language models, and get stable improvement on news recommendation task.
However, all the above methods train models based on centralized data, which is highly privacy-sensitive.
Such kind of collections and analysis of private data have led to privacy concerns and risks~\cite{shin2018privacy}.
Besides, the adoption of some data protection regulations, such as GDPR\footnote{https://gdpr-info.eu}, gives news platforms restrictions and high pressure of using user data to prevent user data leakage.
Different from these methods, we do not use centralized storage for training in our framework, which can better preserve user privacy.

\subsection{Federated Learning}
Federated Learning~\cite{mcmahan2017communication} is an effective method for privacy-preserving model training.
It enables several users to collaboratively train models without sharing their data to a central server.
In federated learning, users first request the latest updated model from central server, and compute local gradients with their local private data.
Central server aggregates the gradients to update the global model and distributes the updated global model to user local devices.
Since the local gradients may leak some private information of users~\cite{bhowmick2018protection, melis2019exploiting}, several privacy protection methods are applied,
such as secure multi-party computation (MPC)~\cite{crypten2020}, differential privacy (DP)~\cite{ren2018textsf}, and homomorphic encryption (HE)~\cite{aono2017privacy}.

Recently, several works have proposed to leverage federated learning in recommendation scenario.
~\citet{Ammaduddin2019FederatedCF} propose federated collaborative filtering (FCF).
In FCF, users use their private rate data to compute gradients of user embeddings and item embeddings.
The user embeddings are updated locally by the gradients of user embeddings, and the gradients of item embeddings are aggregated to update global item embeddings.
~\citet{chai2020secure} propose secure federated matrix factorization (FMF).
FMF is similar to FCF but updates user embeddings and item embeddings according to matrix factorization algorithm.
However, FCF and FMF are not suitable for news recommendation scenarios, since they represent items with ID embeddings and there is much fresh news generated every day.
~\citet{qi-etal-2020-privacy} propose a privacy-preserving method for news recommendation model training.
In FedRec, users use their local data to compute gradients of the model parameters.
A group of randomly sampled users sends their local gradients to the central server to update the global model.
However, the communication and computation cost of FedRec is unacceptable for user devices with limited resource due to the large size of news recommendation models, especially their news models.
In this paper, we propose Efficient-FedRec to reduce the overhead on clients.
We decompose the news recommendation model into a large news model maintained in server and a light-weight user model shared between clients and server.
A small number of news representations and user model are communicated.

\section{Methodology}
In this section, we introduce our Efficient-FedRec method for privacy-preserving news recommendation.
We first introduce the problem formulation and news recommendation framework.
Then we introduce the details of our Efficient-FedRec framework.
The details of secure aggregation are demonstrated in the last subsection.

\subsection{Problem Formulation}
~\label{sec:problem}
Denote $\mathcal{U}=\{u_1, u_2, ... u_P\}$ as user set, where $P$ is the user number.
Given a user $u$, his private behaviors $\mathcal{B}_{u}$ are locally stored on his devices.
In our approach, we denote all news in user behaviors of user $u$ as $\mathcal{N}_u$.
The news recommendation model is decomposed into a news model with parameter set $\Theta_n$ and a user model with parameter set $\Theta_u$.
The server maintains the news models and generates news representations with parameter set $\Theta_e$, and keeps a global user encoder.
The goal is to collaboratively train an accurate news recommendation model without leaking users' private information.

\subsection{News Recommendation Framework}
\begin{figure}[!t]
  \centering
  \includegraphics[width=0.45\textwidth]{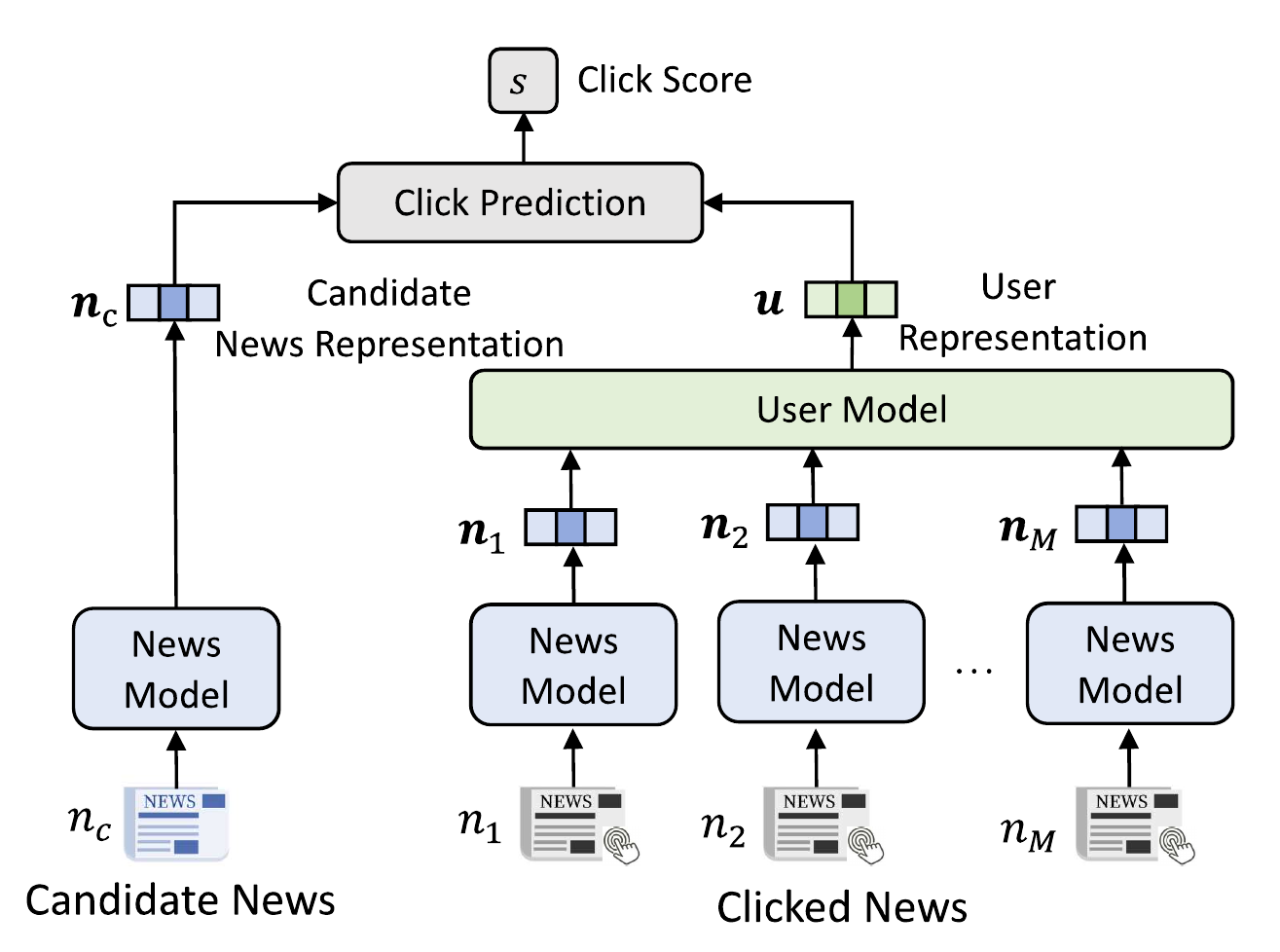}
  \caption{News recommendation framework.}
  \label{fig:comm-newsrec}
\end{figure}
In this subsection, we introduce the news recommendation framework, which is shown in Figure~\ref{fig:comm-newsrec}.
It is composed of two core modules, i.e., news model and user model.

\begin{figure*}[!t]
  \centering
  \includegraphics[width=0.98\textwidth]{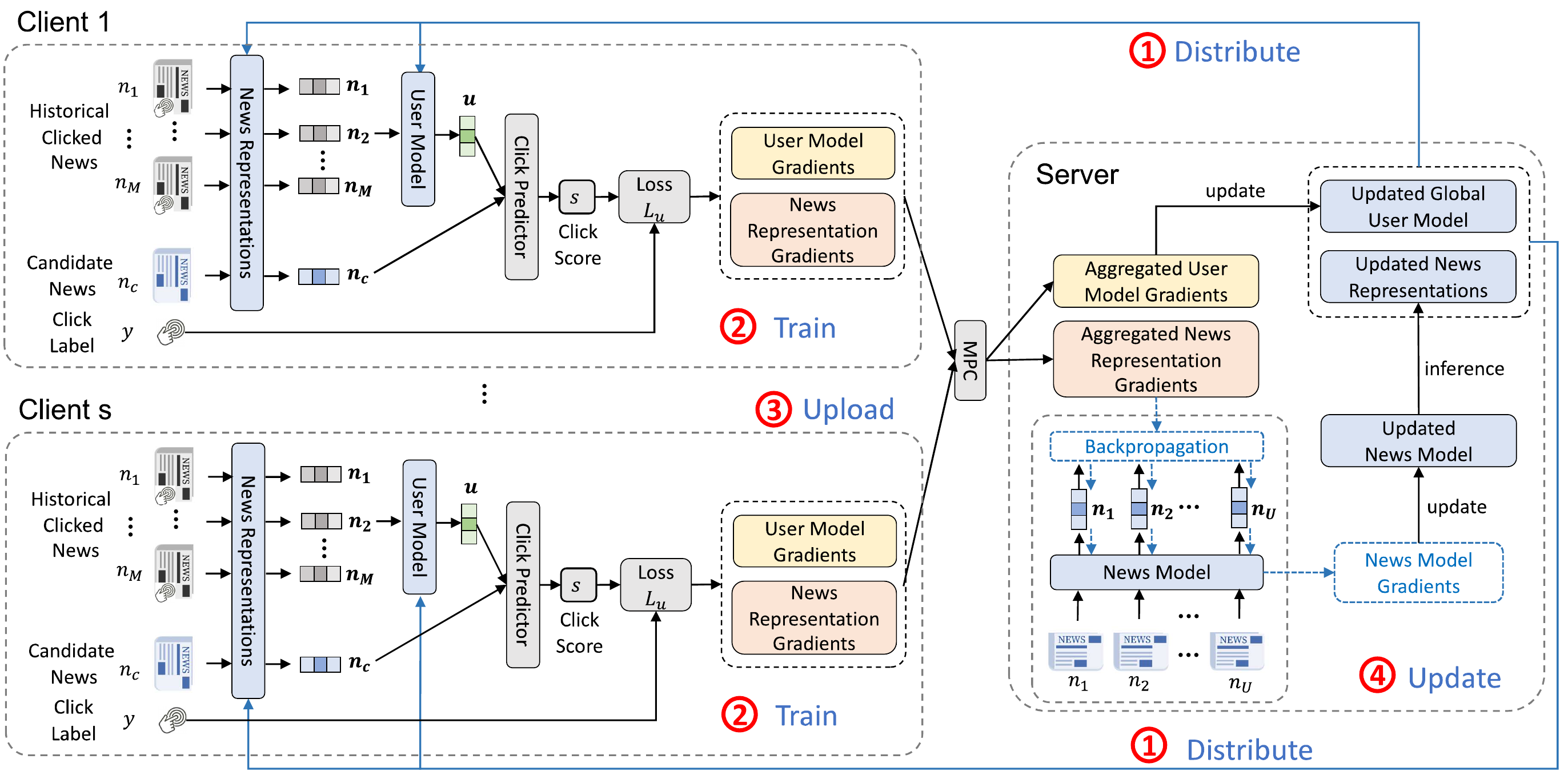}
  \caption{ The framework of Efficient-FedRec.}
  \label{fig:eff-frame}
\end{figure*}

\paragraph{News Model} 
Given a news $n$, the news model is used to learn news representations $\textbf{n}$ from news contents.
It can be implemented by various model structures.
Several existing news recommendation methods use shallow NLP models.
~\citet{wu2019neural} use a combination of multi-head self attention network and additive attention network, ~\citet{an-etal-2019-neural} use a combination of CNN network and additive attention network.
With the success of pre-trained language models (PLM) in NLP, a few methods start to apply pre-trained language models in news model.
~\citet{wu2021empowering} propose PLM-NR, which uses a combination of pre-trained language model and additive attention network as news model.
In our Efficient-FedRec, we apply the news model of PLM-NR~\cite{wu2021empowering}.

\paragraph{User Model}

The user model is used to learn user representations from user historical clicked news.
Denoting the news representations of user historical clicked news $[\textbf{n}_1, \textbf{n}_2, ... \textbf{n}_M]$ as input, the user model computes user representations $\textbf{u}$ as output.
It can be implemented by several model structures.
~\citet{wang2018dkn} use candidate-aware attention, and ~\citet{an-etal-2019-neural} combines user id embeddings and GRU network.
In our Efficient-FedRec, we apply the user model of NRMS~\cite{wu2019neural}, which uses a combination of multi-head self attention network and additive attention network.


\subsection{Framework of Efficient-FedRec}
\label{sec:framework}
In this subsection, we introduce the framework of our Efficient-FedRec.
Each user who participates in model training is called a client.
In our framework, client behaviors are locally stored on their devices, which prevents the risks of data leakage.
Since data of a single user is not enough to train an intelligent news recommendation model, our framework enables multiple clients to collaboratively train a news recommendation model.
To lower the communication and computation overhead on the client side, we decompose the news recommendation model into a large news model maintained on server and a light-weight user model shared on both server and clients.
At the $t$-th round, the model updating contains four steps, i.e., distributing user model and news representations, training local user model and news representations, gradient aggregation and global model updating.
The framework of our Efficient-FedRec is shown in Figure~\ref{fig:eff-frame}.

The first step is distributing user model and news representations.
Since the news model is heavy and users only need the news representations to predict click scores, in our framework users request a small number of news representations in their behaviors and user model instead of the whole model from the central server.
However, directly requesting the news representations of news in user behaviors $\mathcal{N}_u$ will leak user private information.
In our work, we randomly sample a group of clients, who exchange the representations of union news set involved by a group of user behaviors through a secure aggregation protocol (introduced in Section~\ref{sec:mpc}).
Denoting the group of clients as ~$\mathcal{U}_s = \{u_1, u_2 ... u_s\}$, the union news set is computed as $\mathcal{N}_s = \cup_{u_i\in\mathcal{U}_s}\mathcal{N}_i$.
Thus the server only knows the news accessed by a group of clients.
Finally, users keep a local copy of user model $\Theta_u^t$, and news representations of union news set $\Theta_{e_s}^t$.
It is noted news representations of union news set is much smaller than the news model (analyzed in Section~\ref{sec:eff}), which alleviates the communication cost.

The second step is training local user model and news representations.
Given a client $u$, we use his historical clicked news representations to compute user representations $\textbf{u}$ through the local user encoder.
For a candidate news $n_c$, we use the candidate news representation $\textbf{n}_c$ and the user representation $\textbf{u}$ to compute a click score $s$ through a click predictor, which is dot-product in our framework.
Following the previous work~\cite{wu2019neural, an-etal-2019-neural,qi-etal-2020-privacy}, we utilize categorical cross-entropy loss for training.
More specifically, for every clicked candidate news, we sample $K$ non-clicked news in the same impression.
Denote the label of the $i$-th news and user $u$ is $y^i$ and the prediction score is $s^i$, the loss of a training sample is computed as follows:
\begin{equation}
    \mathcal{L}_u^j = -\sum_{i=1}^{K+1}y^i\times log(\frac{exp(s^i)}{\sum_{k=1}^{K+1}exp(s^k)}).
\label{eq:loss}
\end{equation}
The final loss is the average loss of all training samples in $\mathcal{B}_u$, which is computed as follows:
\begin{equation}
    \mathcal{L}_u = -\frac{1}{|\mathcal{B}_u|}\sum_{j=1}^{|\mathcal{B}_u|}\mathcal{L}_u^j
\label{eq:loss_all}
\end{equation}
Denote the local gradients of user encoder as $g_{vu}^t$ and the local gradients of news representations as $g_{e_su}^t$, which are computed as follows:
\begin{equation}
    \begin{aligned}
        &g_{vu}^t = \frac{\partial\mathcal{L}_u}{\partial\Theta_u^t},
        &g_{e_su}^t =\frac{ \partial\mathcal{L}_u}{\partial\Theta_{e_s}^t}.
    \end{aligned}
\end{equation}
In this step, since clients only compute the user model, the computation cost on clients is alleviated.

The third step is gradient aggregation.
The server needs to compute the weighted sum of gradients of user model, news representations, and the sample number from the randomly sampled user group $\mathcal{U}_s$.
Since the local gradients may contain some private information~\cite{bhowmick2018protection, melis2019exploiting}, we apply the secure aggregation to compute the summations (introduced in Section~\ref{sec:mpc}).
The aggregated gradients of user model and news representations are denoted as $\overline{g}_v^t$ and $\overline{g}_{e_s}^t$, which are formulated as follows:
\begin{equation}
    \begin{aligned}
        &\overline{g}_v^t = \frac{1}{\sum_{u \in \mathcal{U}_s}|\mathcal{B}_u|} \sum_{u \in \mathcal{U}_s} |\mathcal{B}_u| \cdot g_{vu}^t, \\
        &\overline{g}_{e_s}^t = \frac{1}{\sum_{u \in \mathcal{U}_s}|\mathcal{B}_u|} \sum_{u \in \mathcal{U}_s} |\mathcal{B}_u| \cdot g_{e_su}^t,
    \end{aligned}
\label{eq:useragg}
\end{equation}
It is noted that each user only sends the gradients of news representations in the union news set $\mathcal{N}_s$, which is much smaller than the news model.
Thus we alleviate the communication overhead.

The final step is global model updating.
The global user model and news model are updated separately.
The global user model is directly updated by gradients of user model through FedAdam algorithm ~\cite{reddi2020adaptive} as follows:
\begin{equation}
    \begin{aligned}
        &\Delta_u^t = \beta_1 \Delta_u^{t-1} + (1-\beta_1) \overline{g}_u^t, \\
        &v_u^t = \beta_2 v_u^{t-1} + (1-\beta_2) {\Delta_u^t}^2, \\
        &\Theta_u^{t+1} = \Theta_u^t + \eta \frac{\Delta_u^t}{\sqrt{v_u^t+\tau}},
    \end{aligned}
\label{eq:userupdate}
\end{equation}
where $\eta$ is the learning rate, $\beta_1$, $\beta_2$ and $\tau$ are parameters of FedAdam.
The news model is updated through a backpropagation training process.
For each news in the union news set $n_i\in \mathcal{N}_s$, the central server has its content and the gradients of its news representation $\overline{g}_{e_i}^{t}\in \overline{g}_{e_s}^{t}$.
We use the news content as input and compute its news representation $\mathbf{n}_i$ through the news model.
The gradients of news model $\overline{g}_n^t$ are computed as follows:
\begin{equation}
    \overline{g}_n^t = \sum_{n_i\in \mathcal{N}_s}\overline{g}_{e_i}^t \cdot \frac{\partial\textbf{n}_i}{\partial\Theta_n^t},
\label{eq:newsgrad}
\end{equation}
where $\Theta_n^t$ is the parameters of news model at the $t$-th round.
We use Adam optimizer to updated new model, which in computed as follows:
\begin{equation}
    \begin{aligned}
        &\Delta_n^t = \beta_1 \Delta_n^{t-1} + (1-\beta_1) \overline{g}_n^t, \\
        &v_n^t = \beta_2 v_n^{t-1} + (1-\beta_2) {\Delta_n^t}^2, \\
        &\Theta_n^{t+1} = \Theta_n^t + \eta \frac{\Delta_n^t}{\sqrt{v_n^t+\tau}},
    \end{aligned}
\label{eq:newsupdate}
\end{equation}
where $\eta$ is the learning rate, $\beta_1$, $\beta_2$ and $\tau$ are hyper parameters of Adam.
We further use the updated news model to infer news representations.
Finally, the updated news representations and user encoder are distributed to all clients.

\subsection{Secure Aggregation}
~\label{sec:mpc}
In this subsection, we first introduce secure aggregation proposed by \citet{bonawitz2017practical}, and then introduce how we apply it to our framework for secure gradients aggregation and news representations distributing.
The secure aggregation is mainly based on multi-party computation (MPC).
It aims to let central server compute weighted sum of vectors without accessing the local vectors of each client in federated learning scenario.
Denoted the local vectors of clients as $\{\textbf{v}_1, \textbf{v}_2, ... \textbf{v}_n\}$, the secure aggregation computes $\textbf{v}=\sum_{i=1}^n\textbf{v}_i$ in a privacy-preserving way.
Meanwhile, it solves the user drop problem on mobile devices.

As we introduce in Section~\ref{sec:framework}, we use the secure aggregation twice.
The first time is to compute the union news set $\mathcal{N}_s$ of a group of users.
Given a user $u_i$, we first transform his local news set $\mathcal{N}_i$ into a local vector $\textbf{h}_i$, of which dimension equals the number of all news.
The $\textbf{h}_i$ is defined as follows:
\begin{equation}
    \textbf{h}_i^j = \left\{
        \begin{aligned}
            &random, &if\ n_j \in \mathcal{N}_i \\
            &0, &otherwise\\ 
        \end{aligned}
    \right.
\label{eq:trans}
\end{equation}
where $\textbf{h}_i^j$ is the $j$-th dimension of $\textbf{h}_i$, and $n_j$ is the $j$-th news in the total news set.
We apply secure aggregation to compute the sum of vectors $\textbf{h} = \sum_{u_i\in\mathcal{U}_s}\textbf{h}_i$.
The inverse transformation of Eq~\ref{eq:trans} is used to compute the union news set $\mathcal{N}_s$ from $\textbf{h}$.
The sampled group of users then request the news representations in the union news set $\mathcal{N}_s$ from central server.

The second time is to securely aggregate gradients.
Each user flattens their local weighted gradients of news representations $|\mathcal{B}_u|\cdot g_{e_su}^t$, local gradients of user model $|\mathcal{B}_u|\cdot g_{vu}^t$ and their sample number $|\mathcal{B}_u|$ to a vector, and applies secure aggregation to compute the summation.
It is noted that only the news in the union news set has the gradients of news representations.
\section{Experiments}
In this section, we demonstrate the efficiency and effectiveness of our Efficient-FedRec.
We conduct several experiments to answer the following research questions:
\begin{itemize}
    \item \textbf{RQ1:} How does our method perform compared with baseline methods?
    \item \textbf{RQ2:} Are the communication and computation overhead significantly reduced compared with baseline methods?
    \item \textbf{RQ3:} How does the news model size influence the performance and overhead of our framework?
    \item \textbf{RQ4:} How does the user group size influence the risk of user information leakage and the effectiveness of our method?
    \item \textbf{RQ5:} How does the user number influence the performance of our framework?
\end{itemize}
\begin{table}[!t]
\centering
\scalebox{0.85}{
\begin{tabular}{c|cc}
\Xhline{1.5pt}
                     & MIND          & Adressa          \\ \hline
\#news               & 65,238              & 20,428        \\
\#users              & 94,057              & 640,503         \\
\#impressions        & 230,117             & -        \\
\#positive samples & 347,727             & 3,101,991        \\
\#negative samples & 8,236,715           & -      \\
\Xhline{1.5pt}
\end{tabular}
}
\caption{Statistics of MIND and Adressa datasets.}
\label{tab:stat}
\end{table}
\subsection{Dataset and Experimental Settings}
We conduct thorough experiments on two public datasets, i.e., MIND\footnote{https://msnews.github.io/} and Adressa\footnote{http://reclab.idi.ntnu.no/dataset/}.
MIND\footnote{We use the small version of MIND for fast experiments.} ~\cite{wu-etal-2020-mind} is a public dataset collected on Microsoft News website in six weeks.
Adressa~\cite{10.1145/3106426.3109436} is publicly released by Adresseavisen,
a local newspaper company in Norway.
Following ~\cite{qi-etal-2020-privacy} and ~\cite{hu-etal-2020-graph}, we use the 6-th day's click to build training dataset and construct historical clicks from the first 5 days' samples.
We randomly sample 20\% clicks from the last day's clicks for validation and the rest clicks for testing.
The historical clicks of validation and testing dataset are constructed from the first 6 days' samples.
Since Adressa does not contain negative samples, we randomly sample 20 news for each click for testing.
The detailed dataset statistics are summarized in Table~\ref{tab:stat}.
Following many previous news recommendation works ~\cite{wu-etal-2020-mind, an-etal-2019-neural,qi-etal-2020-privacy,ijcai2021-224,ijcai2020-418}, we use AUC, MRR, nDCG@5 and nDCG@10 as evaluation metrics.

In our experiments, we apply BERT-Base~\cite{Devlin2019BERTPO} for MIND and nb-bert-base~\cite{kummervold-etal-2021-operationalizing} for Adressa to initialize the pre-trained language model in news encoder.
The dimension of news representations is 400.
To mitigate overfitting, we apply dropout in user model.
The dropout rate is 0.2.
The learning rate is 0.00005.
The number of negative samples associated with each positive sample is 4.
The user group size is 50 on both MIND and Adressa.
All hyper-parameters are selected according to results on the validation set.
We repeat each experiment 5 times independently, and report the average results with standard deviations.

\subsection{Performance Evaluation (RQ1)}
\begin{table*}[]
\centering
\scalebox{0.75}{
\begin{tabular}{c|cccc|cccc}
\Xhline{1.5pt}
\hline
\multirow{2}{*}{Method} & \multicolumn{4}{c|}{MIND} & \multicolumn{4}{c}{Adressa} \\ \cline{2-9} 
                        & AUC       & MRR       & nDCG@5       & nDCG@10      & AUC       & MRR       & nDCG@5       & nDCG@10 \\ \hline
DFM                     & 60.67$\pm$0.20& 28.08$\pm$0.13&    29.93$\pm$0.13&    35.68$\pm$0.13 & 59.90$\pm$1.20& 32.68$\pm$0.75& 29.69$\pm$0.93   & 36.43$\pm$1.11\\ \hline
DKN                     & 64.72$\pm$0.19& 30.53$\pm$0.13&    33.01$\pm$0.15&    38.70$\pm$0.16 & 73.73$\pm$0.48& 39.52$\pm$1.34& 40.98$\pm$1.24   & 47.48$\pm$0.86\\
LSTUR                   & 66.90$\pm$0.08& 32.45$\pm$0.07&    35.11$\pm$0.07&    40.82$\pm$0.07 & 68.37$\pm$2.63& 38.76$\pm$2.14& 38.11$\pm$2.39   & 44.33$\pm$2.42\\
NAML                    & 66.10$\pm$0.25& 31.91$\pm$0.23&    34.52$\pm$0.26&    40.21$\pm$0.24 & 73.09$\pm$1.53& 44.27$\pm$1.53& 43.51$\pm$1.89   & 50.02$\pm$1.71\\
NRMS                    & 66.67$\pm$0.21& 32.25$\pm$0.09&    34.88$\pm$0.11&    40.74$\pm$0.11 & 75.31$\pm$0.94& 42.24$\pm$0.92& 44.66$\pm$1.50   & 48.46$\pm$1.19\\ 
CenRec                  & 66.92$\pm$0.17& 32.30$\pm$0.11&    35.05$\pm$0.13&    40.78$\pm$0.14 & 72.85$\pm$1.53& 40.82$\pm$1.73& 41.62$\pm$2.24   & 47.54$\pm$1.47\\
PLM-NR                  & 67.79$\pm$0.29& 33.16$\pm$0.18&    36.08$\pm$0.21&    41.81$\pm$0.21 & 78.20$\pm$1.28& 47.26$\pm$1.73& 48.41$\pm$2.10   & 54.60$\pm$1.64\\ \hline
FCF                     & 50.02$\pm$0.24& 22.37$\pm$0.18&    22.77$\pm$0.17&    29.02$\pm$0.17 & 51.39$\pm$0.74& 18.98$\pm$1.57& 15.42$\pm$1.72   & 22.94$\pm$1.30\\
FedRec                  & 66.54$\pm$0.18& 31.96$\pm$0.07&    34.54$\pm$0.09&    40.30$\pm$0.09 & 71.73$\pm$1.72& 41.37$\pm$2.21& 41.81$\pm$2.35   & 47.18$\pm$2.09\\
FedRec(BERT)            & 67.45$\pm$0.10& 32.80$\pm$0.10&    35.44$\pm$0.16&    41.35$\pm$0.14 & 78.60$\pm$1.82& 43.81$\pm$0.95& 45.76$\pm$0.89   & 52.64$\pm$1.68\\ \hline
Efficient-FedRec        & 67.44$\pm$0.20& 32.79$\pm$0.06&    35.62$\pm$0.06&    41.35$\pm$0.07 & 79.08$\pm$1.18& 45.09$\pm$1.87& 47.13$\pm$2.35   & 53.85$\pm$1.69\\ 
\Xhline{1.5pt}
\end{tabular}
}
\caption{Results of different news recommendation methods.}
\label{tab:performance}
\end{table*}
In this section, we compare our Efficient-FedRec framework for privacy-preserving news recommendation with several baseline methods, including
news recommendation methods with centralized storage:
(1) DFM~\cite{lian2018towards}, a multi-channel deep fusion model for news recommendation;
(2) DKN~\cite{wang2018dkn}, a knowledge-aware news recommendation method;
(3) LSTUR~\cite{an-etal-2019-neural}, using user id embedding to capture user long-term interests, and GRU network to capture short-term interests;
(4) NAML ~\cite{wu-etal-2019-neural-news-recommendation}, learning news representations via multi-view learning;
(5) NRMS ~\cite{wu2019neural}, using two self-attention networks for better news and user modeling;
(6) CenRec ~\cite{qi-etal-2020-privacy}, a central version of FedRec;
(7) PLM-NR ~\cite{wu2021empowering}, applying pre-trained language model to empower the performance of news recommendation. For fair comparison, we use the user model in NRMS.
privacy-preserving news recommendation methods:
(8) FCF ~\cite{Ammaduddin2019FederatedCF}, federated collaborative filtering for recommendation;
(9) FedRec ~\cite{qi-etal-2020-privacy}, privacy-preserving method for news recommendation model training. For fair comparison, we do not add differential privacy;
(10) FedRec(BERT), applying FedRec to train PLM-NR in a privacy-preserving way.
our method:
(11) Efficient-FedRec, using our Efficient-FedRec framework to train PLM-NR in a privacy-preserving and efficient way.
The experimental results of all these methods are shown in Table~\ref{tab:performance}.

We have several observations from Table~\ref{tab:performance}.
First, comparing our Efficient-FedRec with SOTA news recommendation methods with centralized storage (DKN, NAML, NRMS, LSTUR and PLM-NR), our method achieves comparable performance.
Moreover, our method does not need users to share their behavior data.
Therefore, it validates our method can train accurate news recommendation models and meanwhile protect user privacy.
Second, our method performs better than FCF.
This is because FCF is not suitable for news recommendation, since there are severe cold-start problems in news recommendation scenario~\cite{qi-etal-2020-privacy, wu-etal-2020-mind}.
Third, our Efficient-FedRec outperforms FedRec.
This is because we use pre-trained language model in news model, which can help better understand the semantics of news contents.
Forth, comparing our Efficient-FedRec with FedRec(BERT), our Efficient-FedRec achieves comparable performance.
This is because our method has the same gradients as FedRec(BERT) if dropout and batch normalization are not applied in news model.
Finally, FedRec(BERT) and Efficient-FedRec perform worse than PLM-NR, and FedRec performs worse than CenRec.
This is probably because user behaviors are non-i.i.d, which may make it difficult for federated learning to achieve good results~\cite{mcmahan2017communication,wang2019adaptive}.
\begin{table*}[!t]
\centering
\scalebox{0.9}{
\begin{tabular}{c|c|ccc|ccc}
\Xhline{1.5pt}
\multirow{3}{*}{BERT} & \multirow{3}{*}{AUC} & \multicolumn{3}{c|}{Efficient-FedRec} & \multicolumn{3}{c}{FedRec}    \\ \cline{3-8} 
                      &                      & Comm.    & Comp.    & Comp.    & Comm.    & Comp.    & Comp.    \\
                      &                      & Cost     & Cost     & Cost     & Cost     & Cost     & Cost     \\
                      &                      & (client) & (client) & (server) & (client) & (client) & (server) \\ \hline
Tiny                  & 64.21                & 2.18M    & 0.02s    & 2.05s    & 10.01M   & 0.69s    & 0.01s    \\
Mini                  & 65.55                & 2.18M    & 0.02s    & 3.20s    & 23.74M   & 2.44s   & 0.01s    \\
Small                 & 65.92                & 2.18M    & 0.02s    & 5.88s    & 59.32M   & 9.03s   & 0.01s    \\
Medium                & 67.05                & 2.18M    & 0.02s    & 6.39s    & 84.54M   & 19.55s   & 0.01s    \\
Base                  & 67.44                & 2.18M    & 0.02s    & 6.74s    & 221.29M  & 51.92s  & 0.02s    \\
Large                 & 67.50                & 2.18M    & 0.02s    & 8.81s    & 673.28M  & 117.04s  & 0.04s     \\
\Xhline{1.5pt}
\end{tabular}
}
\caption{Results of different news models on MIND.}
\label{tab:news-model}
\end{table*}
\subsection{Efficiency Analysis (RQ2)}
~\label{sec:eff}
\begin{figure}[!t]
  \centering
  \includegraphics[width=0.45\textwidth]{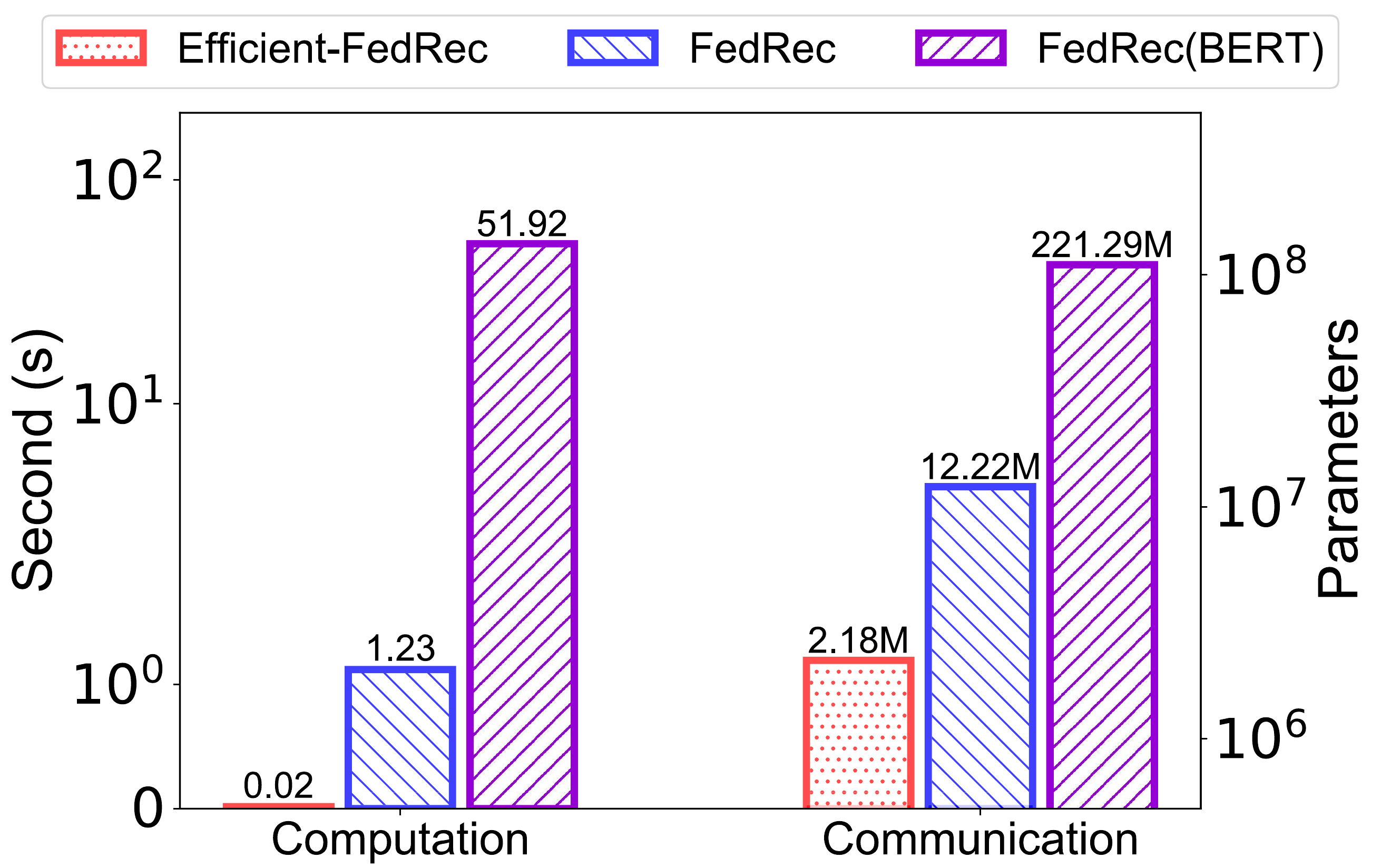}
  \caption{The communication cost and computation cost of privacy-preserving methods on MIND.}
  \label{fig:cost}
\end{figure}

\begin{figure*}
     \centering
     \begin{subfigure}[b]{0.23\textwidth}
         \centering
         \includegraphics[width=\textwidth]{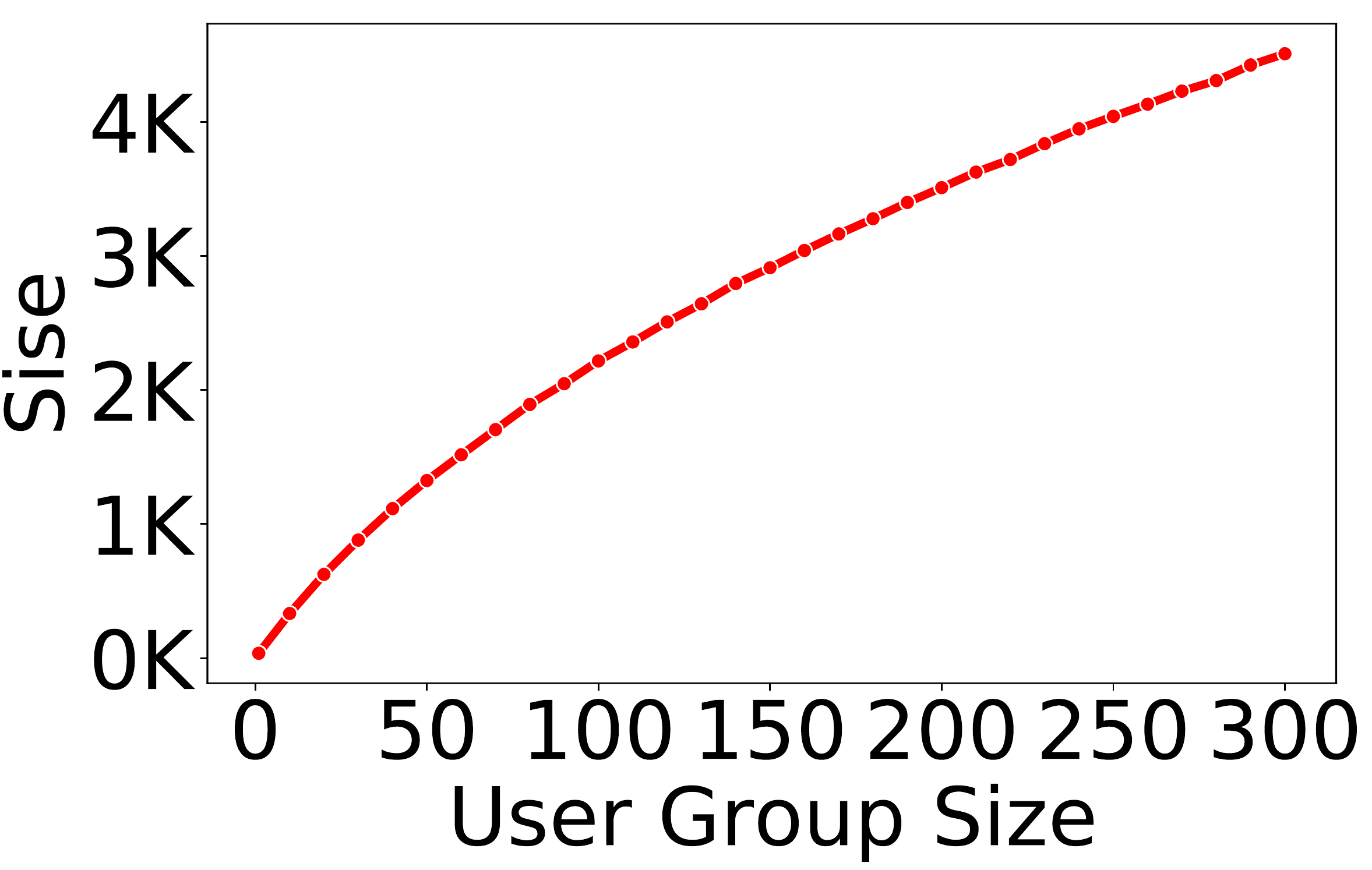}
         \caption{Union news set size.}
         \label{fig:union-news}
     \end{subfigure}
     \hfill
     \begin{subfigure}[b]{0.23\textwidth}
         \centering
         \includegraphics[width=\textwidth]{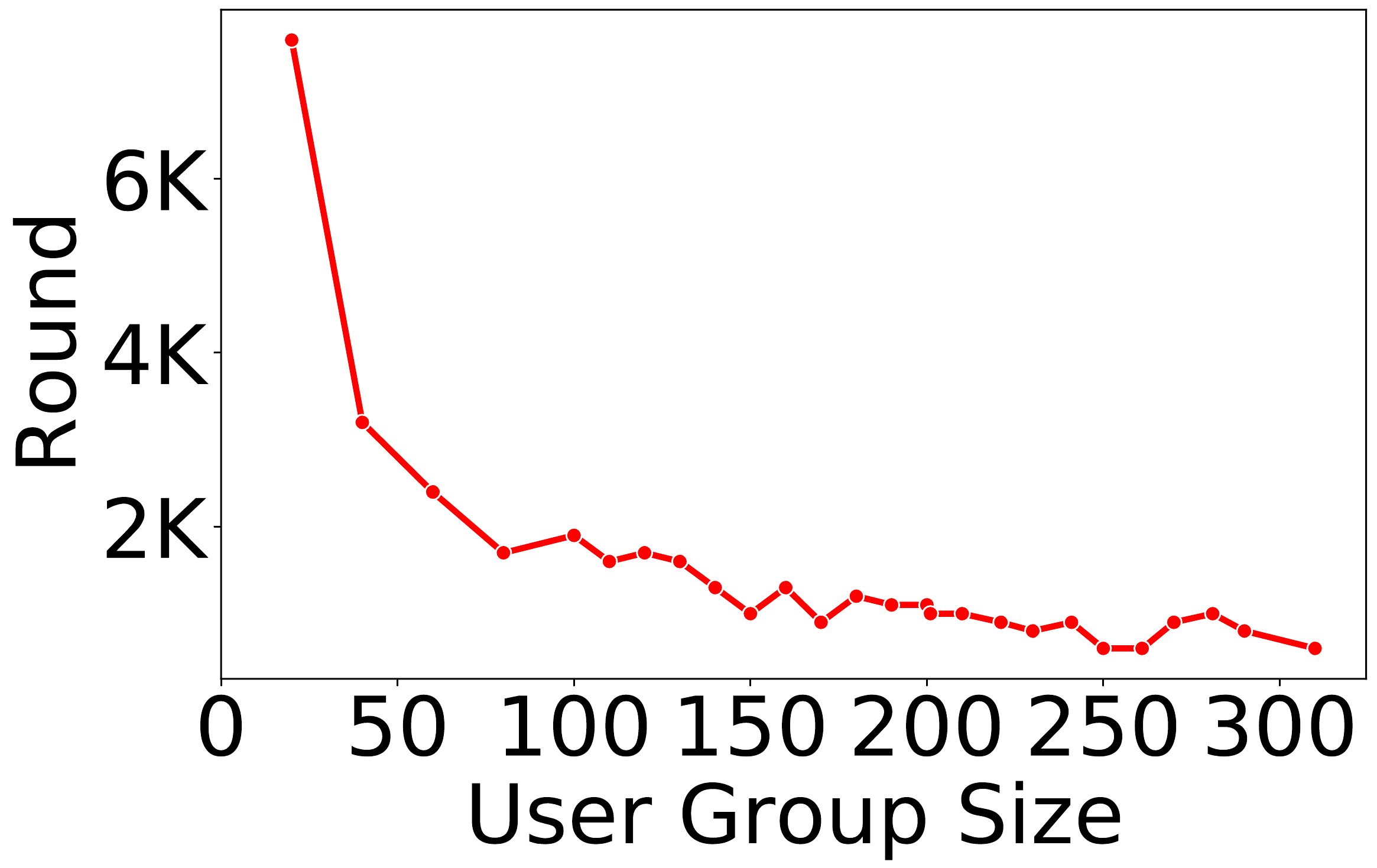}
         \caption{Convergence round.}
         \label{fig:conv-round}
     \end{subfigure}
     \hfill
     \begin{subfigure}[b]{0.25\textwidth}
         \centering
         \includegraphics[width=\textwidth]{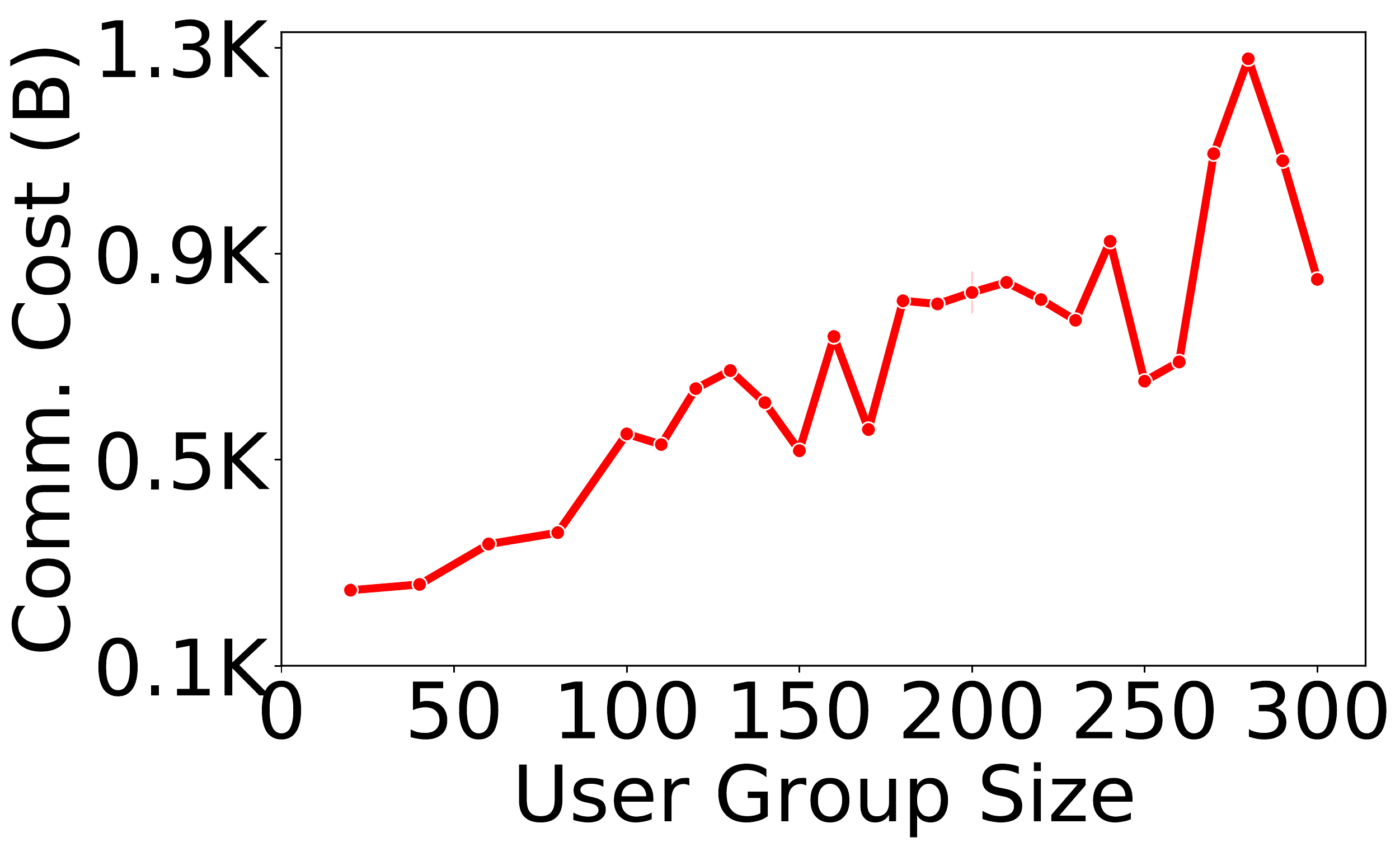}
         \caption{Overall comm. cost.}
         \label{fig:all-comm-cost}
     \end{subfigure}
     \begin{subfigure}[b]{0.23\textwidth}
         \centering
         \includegraphics[width=\textwidth]{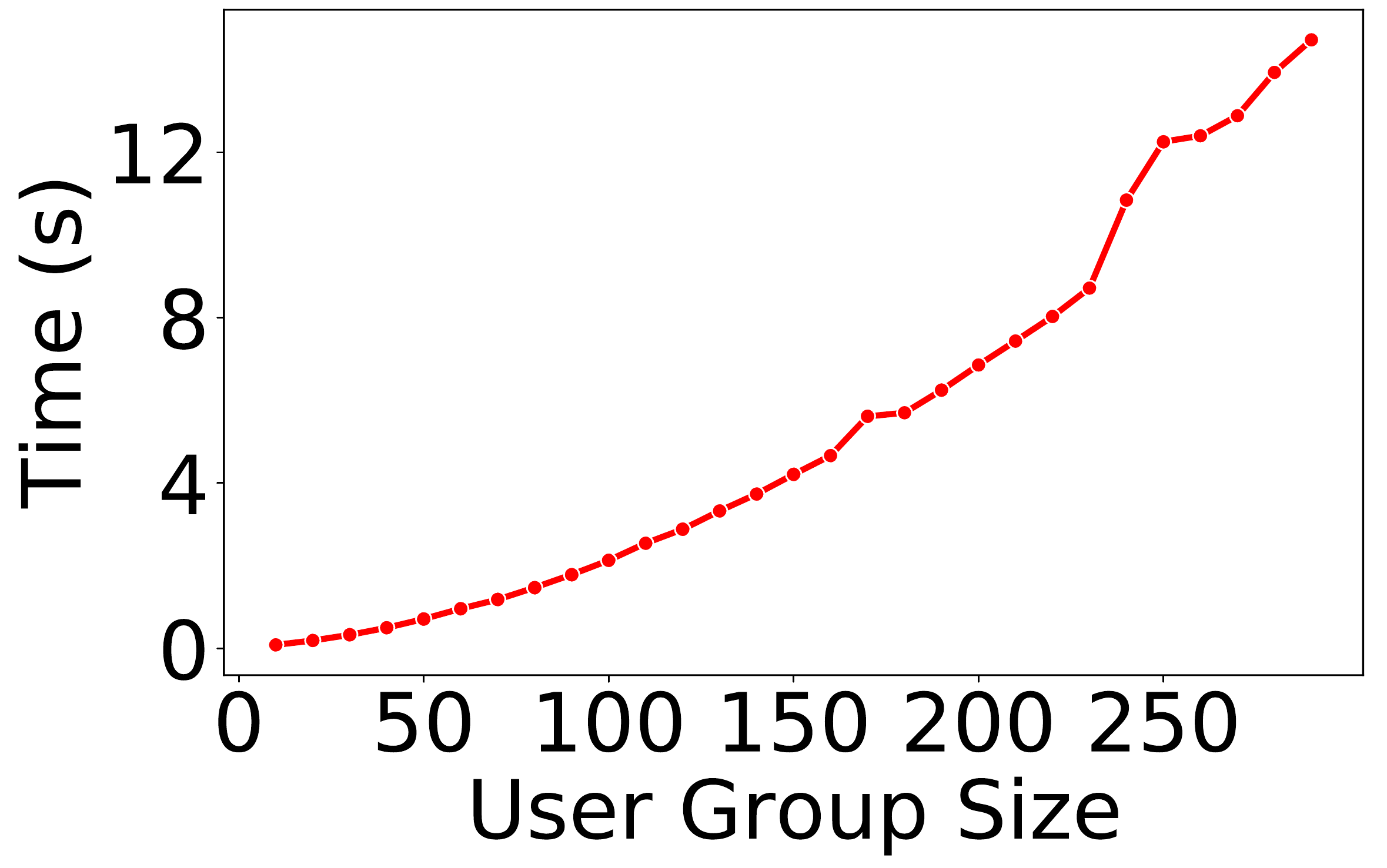}
         \caption{Secure aggregation time.}
         \label{fig:secure-agg}
     \end{subfigure}
    \caption{Impact of user group size on MIND.}
    \label{fig:user-group-size}
\end{figure*}

In this subsection, we analyze the communication and computation cost of our Efficient-FedRec on MIND.
The average size of the union news set is 1,320 per round, the gradient and parameter size of which is 1.06M.
We assume users leverage CPU for calculation.
Figure~\ref{fig:cost} shows the average computation time and the communication overhead of each user per round of several privacy-preserving methods.
From Figure~\ref{fig:cost}, we have several observations. 
First, the average computation time of Efficient-FedRec is lower than those of FedRec and FedRec(BERT).
This is because in our framework users do not need to compute the news model, which lowers the computation overhead.
Second, the communication overhead of Efficient-FedRec is much lower than the overhead of FedRec and FedRec(BERT).
This is because in our framework users request and send the gradients and parameters of user model and a small number of news representations, which is much smaller than the gradients and parameters of the whole model.

\subsection{The Influence of News Model Size (RQ3)}
In this subsection, we apply different size of BERTs in news model to study the influence of the news model size on MIND.
The computation cost of clients is tested on CPU, while the computation cost of server is tested on GPU, which is reasonable since clients are usually with limited computation resource.
The result are shown in Table~\ref{tab:news-model}, where we have several observations.
First, the recommendation performance increases with the news model size, which shows the effectiveness of applying large news model.
Second, the communication and computation cost of our method on clients are lower than FedRec.
This is because in Efficient-FedRec clients only compute the user model and  request the user model and the representations of news involved in their local behaviors.
Additionally, the gap of the overhead between Efficient-FedRec and FeRec becomes larger with larger news model, which demonstrates the superiority of our method in using large news models.
Third, the computation overhead of our method on server is larger than FedRec.
It is because in our framework the news model is trained on central server.
However, the overall computation time of Efficient-FedRec is lower than FedRec.
This is because the server can use powerful GPU clusters to update the news model.
It is noted that we simulate client computation cost with 100\% CPU utilization.
The computation time on real-time devices will be larger than the results reported in Table~\ref{tab:news-model}.

\subsection{Influence of User Group Size (RQ4)}
\label{sec:user-group-size}

In this section, we study the influence of user group size on union news set size, convergence round, overall communication cost and secure aggregation time.
The results are shown in Figure~\ref{fig:user-group-size}.
As shown in Figure~\ref{fig:union-news}, with the increasing of user group size, the size of union news set increases.
When user group size is 40, the average size of union news set is 1,115, which is 10 times larger than the average size of user local news set, i.e., 114.
Therefore, when user group size is large enough, it is hard for server to recover interacted news of users.
Then, we study the impact of user group size on communication cost.
Since larger user group size leads to larger union news set size, the communication cost of per user increases.
In Figure~\ref{fig:conv-round} we also find larger user sizes can make the model converge faster.
The impact of user group size on overall communication cost is shown in Figure~\ref{fig:all-comm-cost}, which is influenced by group user size, communication cost per user and convergence round.
It is shown the overall communication cost increases with larger user group size.
Finally, we study the impact of user group size on secure aggregation time (per user).
As shown in Figure~\ref{fig:secure-agg}, the computation cost of secure aggregation increases with larger user group size.
The computation time of secure aggregation is 0.71s when user group size is 50.
Considering privacy protection ability, communication cost and secure aggregation cost, we set user group size as 50 in our experiment on MIND.

\subsection{Influence of User Number (RQ5)}
\begin{figure}[!t]
  \centering
  \includegraphics[width=0.45\textwidth]{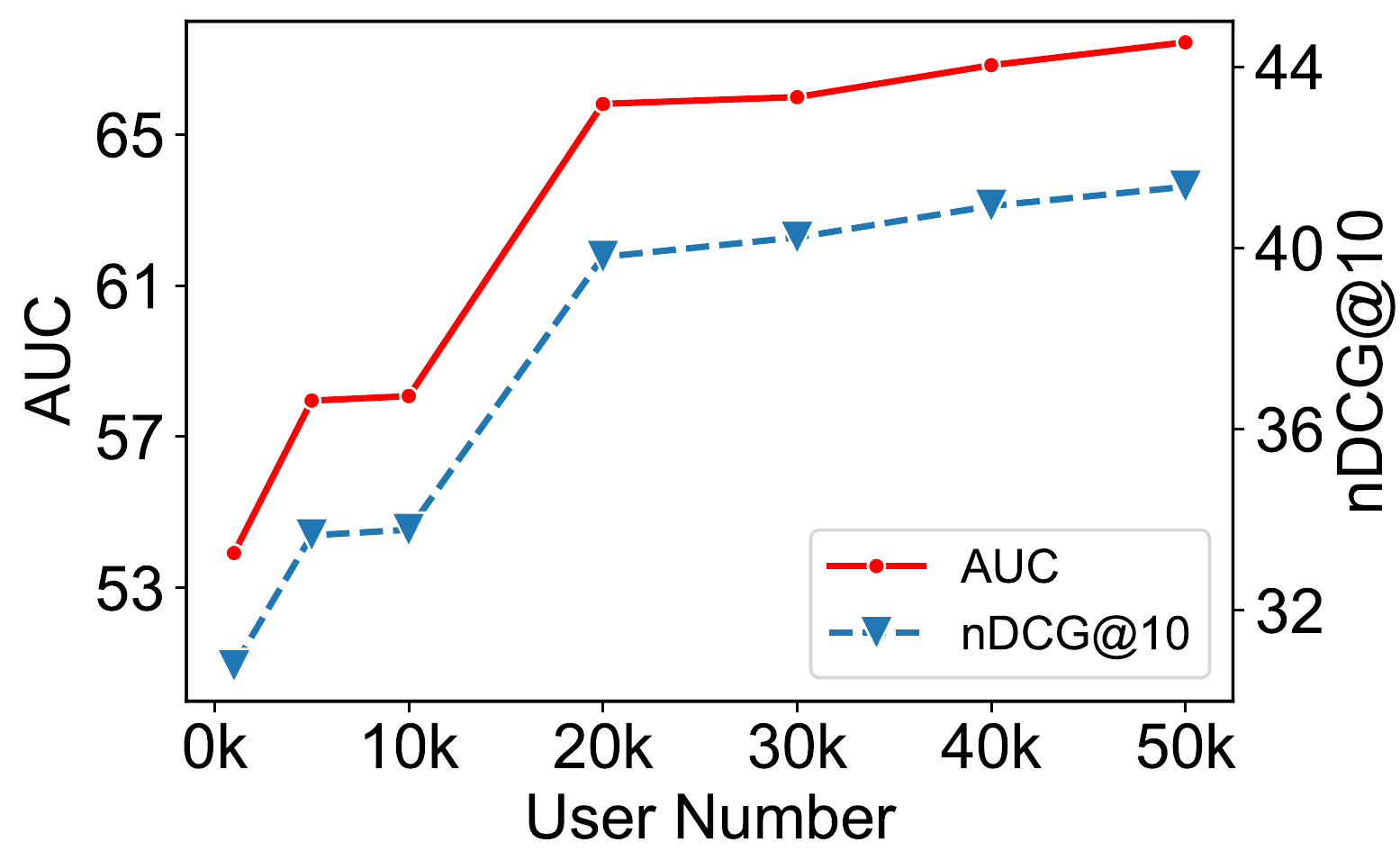}
  \caption{Results of different user numbers on MIND.}
  \label{fig:user}
\end{figure}

In this subsection, we study the influence of the number of users who participate in model training.
We randomly sample different numbers of users from MIND.
The experimental results are shown in Figure~\ref{fig:user}.
We can observe the performance increases with higher user numbers, which validates the idea of training news recommendation collaboratively with a large size of users.
Moreover, it shows our Efficient-FedRec can effectively explore useful information from multiple user behaviors.



\section{Conclusion}
In this paper, we propose an efficient federated learning framework for privacy-preserving news recommendation named Efficient-FedRec.
We decompose the news recommendation model into a large news model maintained by server and a light-weight user model.
Users request news representations and user model from the central server and compute gradients with user local data.
The central server aggregates gradients to update the user model and news model.
The updated news model is further used to infer news representation by server.
In order to protect the private information in user local gradients, we apply secure aggregation to aggregate gradients.
In order to protect user interacted news history, we exchange the news representations in the union news set involved by a group of user behaviors.
Experiments on two real-world datasets validate our method can effectively reduce both communication and computation cost on user side while keep the model performance.
\section*{Ethical Statements}

\textbf{User Information Protection in Dataset}
In this paper, we conduct experiments on two public datasets, i.e., MIND and Adressa.
MIND dataset was released in~\cite{wu-etal-2020-mind}.
It is a public English news recommendation dataset.
In this dataset each user was de-linked from the production system when securely hashed into an anonymized ID using onetime salt mapping to protect user privacy.
We have agreed with Microsoft Research License Terms\footnote{https://msnews.github.io/} before downloading this dataset and complied with these license terms when using this dataset.
Adressa dataset was released in~\cite{10.1145/3106426.3109436}.
It is a public Norwegian news recommendation dataset.
The users in this dataset are anonymized to protect user privacy. 
We follow the dataset license\footnote{https://creativecommons.org/licenses/by-nc-sa/4.0/} when using this dataset.
Thus, all the datasets used in our paper are public datasets where user privacy information is well protected.

\textbf{Influence of User Group Size}
The user groups consist of randomly sampled users in each round to update model according to our framework.
We conduct experiments to analyze the influence of user group size, and the results are summarized in Section~\ref{sec:user-group-size}.
The experimental results show that as long as the user group size is properly large, which is usually easy to satisfy in practical applications, the information of user interacted news can be well protected.

\section*{Acknowledgments}

This work is supported by Youth Innovation Promotion Association of CAS.
We would like to thank Tao Qi and Hao Wang for their great comments and help on experiments.
\bibliography{anthology,custom}
\bibliographystyle{acl_natbib}

\appendix
\clearpage
\section*{Appendix}
~\label{sec:appendix}

\subsection*{Secure Aggregation}
The secure aggregation is mainly based on secure multi-party computation.
It applies secret sharing, pseudorandom generator and key agreement to compute weighted summation of vectors in the federated learning scenario.
Meanwhile, it solves the user drop problem on mobile devices.
The secure aggregation contains four steps, which are key initialization, generating secret shares, computing masked vectors, and unmasking vectors.
Since there is no efficient implementation of secure aggregation, we test the efficiency using our implementation.
The detailed algorithms and hyper-parameters of secure aggregation are listed in Table~\ref{tab:sa}.
\begin{table}[!h]
\centering
\scalebox{0.8}{
\begin{tabular}{ccc}
\Xhline{1.5pt}
             & Algoritm         & Size/Value \\ \hline
Hash         & MD5              & 32B  \\
Signing      & ed25519          & 32B  \\
Key Exchange & x25519           & 32B  \\
Random Seed  & -                & 16B  \\
PRG          & Mersenne Twister & -    \\ 
t of secrete sharing & -      & 25   \\
n of secrete sharing & -      & 50   \\
\Xhline{1.5pt}
\end{tabular}
}
\caption{Algorithms and settings of secure aggregation.}
\label{tab:sa}
\end{table}

\subsection*{Hyper-parameter Settings}
The complete hyper-parameter settings are listed in Table~\ref{tab:hyper}.
\begin{table}[h]
\centering
\scalebox{0.85}{
\begin{tabular}{c|c|c}
\Xhline{1.5pt}
Hyperparameters         & MIND       & Adressa  \\ \hline
learning rate           & 0.00005    & 0.00005\\
negative sampling ratio & 4          & 4      \\
user group size         & 50         & 50  \\
news dimention          & 400        & 400    \\
dropout rate            & 0.2        & 0.2    \\
Adam beta1              & 0.9        & 0.9    \\
Adam beta12             & 0.99       & 0.99   \\
Adam esp                & $10^{-8}$  &$10^{-8}$\\ 
\Xhline{1.5pt}
\end{tabular}
}
\caption{Hyper-parameter settings.}
\label{tab:hyper}
\end{table}

\subsection*{Experimental Environment}
We conduct experiments on a linux server with Ubuntu 18.04.5.
The server has 4 A100-PCIE-40GB with CUDA 11.1.
And the CPU is Intel(R) Xeon(R) Gold 6248R CPU @ 3.00GHz.
We use python 3.6.12 and pytorch 1.8.1.
In our experiments, the PLM-NR and FedRec(BERT) are trained with 4 GPUs.
Our EffFedRec and the other methods are trained on single GPU.
\end{document}